\documentclass[aps,prl,twocolumn,showpacs,10pt,superscriptaddress,preprintnumbers,nofootinbib]{revtex4-1}
\usepackage[hyperindex,breaklinks,hyperfootnotes=false,bookmarks=false]%
{hyperref}
\usepackage{graphicx}
\usepackage{amsmath,bm,braket}
\usepackage{color}
\usepackage{multirow}

\allowdisplaybreaks

\begin{document}

\title{Modeling of $t$-channel single top-quark production at the LHC}
\author{Jun Gao}
\email{jung49@sjtu.edu.cn}
\affiliation{INPAC, Shanghai Key Laboratory for Particle Physics and Cosmology,
School of Physics and Astronomy, Shanghai Jiao Tong University, Shanghai 200240, China}
\affiliation{Center for High Energy Physics, Peking University, Beijing 100871, China}
\author{Edmond L. Berger}
\email{berger@anl.gov}
\affiliation{High Energy Physics Division, Argonne National Laboratory, Argonne, Illinois 60439, USA}

\begin{abstract}
We study the modeling of $t$-channel single top-quark
production at Large Hadron Collider (LHC) energies.
We compare predictions at next-to-next-to-leading order in a 5-flavor
scheme to those of next-to-leading order in a 4-flavor scheme,
finding the two schemes agree within a few percent in general for the shape of
kinematic distributions of the top quark.
The predictions in the 5-flavor scheme show strong stability
for both normalization and distributions, and are superior to those of
the 4-flavor scheme at comparable orders.  
We present comparisons of the predictions with LHC data.  
Our findings provide clear theoretical guidance for precision studies
of single top-quark physics at the LHC.

\end{abstract}

\pacs{}
\maketitle

\pagebreak
\newpage
\noindent \textbf{Introduction.}
As the heaviest particle in the standard model (SM), the top quark ($t$) is thought to offer special opportunities to explore 
electroweak symmetry and possible new physics beyond the SM.  Single top quark production at hadron colliders provides a great
opportunity to directly probe the electroweak $Wtb$ vertex, including measurement of the Cabibbo-Kobayashi-Maskawa (CKM) matrix 
element $V_{tb}$.   
In addition, the data can be used to extract the top-quark mass~\cite{Alekhin:2016jjz,Sirunyan:2017huu}
and to constrain the ratio of $u$-quark to $d$-quark parton
distributions~\cite{1404.7116,Alekhin:2015cza,Berger:2016oht,1912.09543}.
Single top-quark production is also sensitive to physics beyond the
SM~\cite{He:1998ie,hep-ph/0007298}, including modified structure of the $Wtb$ vertex,
new gauge bosons, new
heavy quarks, and top-quark flavor-changing neutral currents.  

The $t$-channel production of a single top quark has the largest rate among all single production 
channels at the Large Hadron Collider.  It occurs via electroweak charged-current coupling with
a bottom quark, where the bottom quark arises from gluon splittings.
The production can be calculated either in a factorization scheme based on 4 flavors (4FS) in the 
initial state or in a factorization scheme that also treats the bottom quark as a massless
parton in initial hadrons (5FS).
Critical questions arise on the use and agreement of the two
heavy-quark schemes in single top quark production, with 
initial efforts at understanding made in Refs.~\cite{0903.0005,1203.6393,1711.02568}.
Large theoretical uncertainties in modeling of the
signals and of various measured quantities at the LHC~\cite{1702.02859,Sirunyan:2018rlu,Aaboud:2019pkc} 
must be addressed in view of the unprecedented precision expected in upcoming high luminosity studies at the 
LHC.
One issue is whether the 4FS provides a better description of kinematic distributions than the 5FS.
In this manuscript we address these questions with a detailed comparison
of the next-to-next-to-leading (NNLO) predictions in 5FS to those at 
next-to-leading order (NLO) in 4FS.
We observe excellent agreement between 5FS and 4FS for predictions
of the shapes of kinematic distributions including the transverse momentum
and rapidity of the top quark.
Predictions in the 5FS further exhibit a better convergence and strong
stability against choice of QCD scales, and are superior to predictions
from the 4FS evaluated at comparable orders.
The agreement of the two schemes provides important confidence in
the reliability of higher order QCD predictions at the LHC.

Significant efforts have been made recently to improve the theoretical
description of $t$-channel single top quark production. 
The NLO QCD corrections in the 5-flavor scheme are calculated 
in Refs.~\cite{NUPHA.B435.23,hep-ph/9603265,hep-ph/9705398,hep-ph/9807340,hep-ph/0102126,
hep-ph/0207055,Sullivan:2004ie,hep-ph/0408158,hep-ph/0510224,hep-ph/0504230,1007.0893,
1012.5132,1102.5267,1305.7088,1406.4403,Carrazza:2018mix}.
Further NNLO QCD corrections are reported in
Refs.~\cite{1404.7116,Berger:2016oht,1708.09405}.
The NLO calculation in the 4-flavor scheme is carried out in Ref.~\cite{0903.0005}.  
The NLO electroweak corrections are also calculated~\cite{1907.12586}. 
Soft gluon resummation is considered in
Refs.~\cite{1010.4509,1103.2792,1210.7698,1510.06361,1801.09656,Kidonakis:2019nqa,Cao:2019uor}.
Matching of NLO
calculations to parton showers is done in the framework of POWHEG and MC@NLO
Refs.~\cite{hep-ph/0512250,0907.4076,1207.5391,1603.01178}.

In the remaining paragraphs 
we present our numerical results on inclusive cross sections
and kinematic distributions and comparisons with LHC data.

\noindent \textbf{Total cross sections.}
The NNLO predictions for single top quark production in the 5-flavor
scheme are calculated using phase-space slicing with the $N$-jettiness
variable~\cite{Stewart:2010tn,Boughezal:2015dva, Gaunt:2015pea,Berger:2016inr}
together with the method of ``projection-to-Born'' in Ref.~\cite{1506.02660}.
Details for the NNLO calculation in the 5FS can be found in Ref.~\cite{1708.09405}.
We use the program MCFM~\cite{Campbell:2016jau,Boughezal:2016wmq} to calculate NLO 
predictions for single top quark production in the 4-flavor scheme.
The original calculation was detailed in Ref.~\cite{0903.0005}.
In both calculations, the QCD corrections can be further
factored as from either fermion line with heavy quarks or light quarks
neglecting certain color suppressed contributions~\cite{Assadsolimani:2014oga,Meyer:2016slj},
which are irrelevant for the comparison.

Schematically the difference of 5FS and 4FS can be understood by
taking cross sections at first comparable order, i.e., leading order (LO) in 4FS
and NLO in 5FS as an example
\begin{align}
\label{eq:sch}
\sigma^{\rm LO}_{\rm 4F}\,\,\, &=\alpha_s(\mu)[a_1\ln(m_t^2/m_b^2)+c_1
+d_1(m_b^2/m_t^2)],
\nonumber \\
\sigma^{\rm NLO}_{\rm 5F}&=\alpha_s(\mu)[a_1\ln(m_t^2/m_b^2)+c_1]
\nonumber\\
&+\alpha_s^2(\mu)[a_2\ln^2(\mu^2/m_b^2)+a_3\ln(\mu^2/m_b^2)]
\nonumber \\
	&+\,{\rm higher\quad orders},
\end{align}
where $\mu$ is the factorization scale and $\alpha_s(\mu)$ is the strong
coupling constant; 
$m_t$ and $m_b$ are masses of the top quark and bottom quark respectively.
Coefficients $a_i$, $c_i$, and $d_i$ are independent of the
bottom quark mass.
Calculations in the 4FS are performed order by order in $\alpha_s$
and include exact bottom quark mass dependence like power correction term
$d_1$ in Eq.(\ref{eq:sch}) which is otherwise neglected in 5FS.
We include only the leading power correction term for the purpose of this illustration.
On another hand, calculations in the 5FS resum potential large logarithms
of bottom quark mass due to gluon splitting into bottom quarks in the initial state 
through all orders in $\alpha_s$, as in terms associated with
$a_i$.
The NLO and NNLO predictions have a resummation accuracy of
next-to-leading and next-to-next-to-leading logarithms.

We focus on results for top quark production at 13 TeV though results are
similar for either top anti-quark or top quark production at 8 TeV.
We use CT14 NNLO PDFs~\cite{Dulat:2015mca} of corresponding flavor numbers throughout 
the comparison and a bottom quark mass of 4.75 GeV and a top quark
mass of 172.5 GeV accordingly.
We set the QCD renormalization scale and factorization scale to be the same, unless
otherwise specified, and
choose different values in the comparisons.

\begin{figure}[h!]
	\begin{center}
		\includegraphics[width=0.45\textwidth]{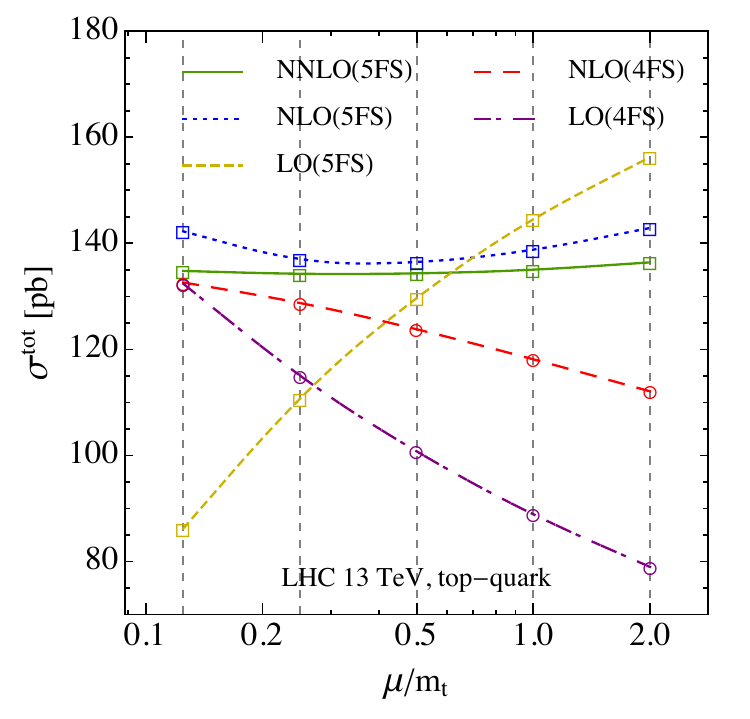}
	\end{center}
	\caption{\label{fig:totinc}
		Inclusive cross sections for single top-quark production at the LHC at 13 TeV 		
		at various orders in QCD, as functions of the renormalization and factorization
		scale in both 5FS and 4FS. 
		}
\end{figure}

In Fig.~\ref{fig:totinc} we plot the total inclusive cross sections for single
top-quark production at 13 TeV as functions of QCD scales.
In 5FS the choice of QCD scale $\mu_{5F}$ determines size of the quasi-collinear
logarithms that are resummed through the bottom quark parton distribution.
Resummation leads to fast convergence of the cross sections and stability
against scale choice at higher orders in 5FS.
For instance the NNLO cross section varies between 134.3 pb to 136.4 pb for
the range of scales considered.
On another hand, predictions in 4FS exhibits larger scale dependence
owing to missing higher order contributions, e.g., with a variation between
112.1 pb to 132.6 pb at NLO.
We note a fair comparison of predictions from the two schemes should be
NNLO(NLO) in 5FS to NLO(LO) in 4FS since contributions from gluon splitting
at large angles are only included starting from NLO in 5FS.
Predictions of the two schemes do approach each other at high orders
as resummed contributions from even higher orders diminish.
From Fig.~\ref{fig:totinc} we conclude a preferable scale choice for
the 5FS of either $\mu_{5F}=m_t/4$ or $m_t/2$ where the NNLO corrections
are small and meanwhile the series show a good convergence, similar to
the case of top quark pair production~\cite{1606.03350}.
Indeed a lower value of the QCD scale in 5FS was suggested
in Ref.~\cite{1203.6393} which shows those quasi-collinear logarithms
to be resummed are accompanied by a universal suppression from phase
space integration.
Unlike the case of 5FS we cannot find a strong motivation for an
optimal scale choice in 4FS though a lower value leads to better
agreement with 5FS on the total cross sections.  
We use a nominal scale of $\mu_{4F}=m_t$ in the following comparisons.

\begin{figure}[h!]
	\begin{center}
		\includegraphics[width=0.45\textwidth]{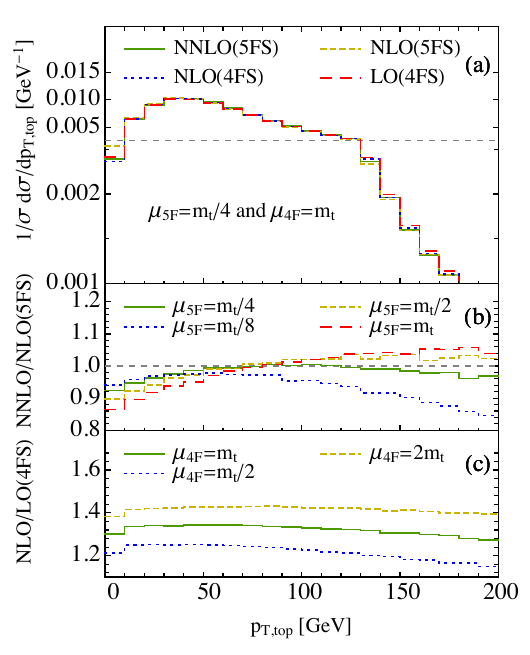}
	\end{center}
	\caption{\label{fig:pt}
		Differential distribution in transverse momentum of a top quark at 13 TeV.
		(a): normalized cross sections with the nominal scale choices for both schemes
		(note below the horizontal dashed line, a different linear scale is used);
		(b) and (c): ratio of NNLO(NLO) to the respective NLO(LO) predictions of absolute
		cross sections with various scale choices in 5FS(4FS).}
\end{figure}

\begin{figure}[h!]
	\begin{center}
		\includegraphics[width=0.45\textwidth]{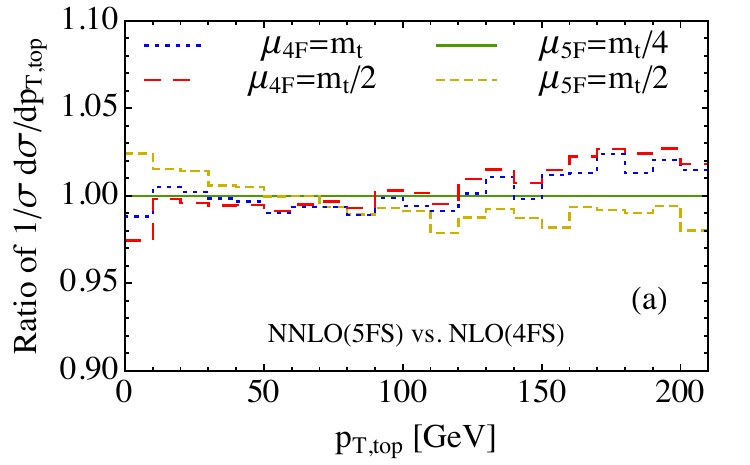}
		\includegraphics[width=0.45\textwidth]{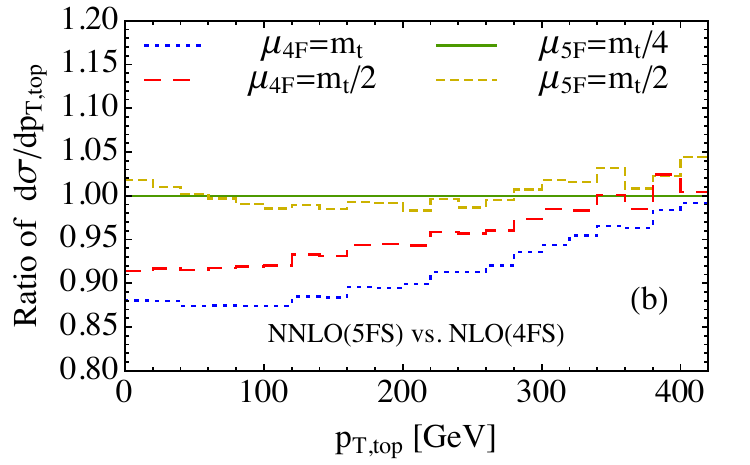}
	\end{center}
	\caption{\label{fig:ptrap}
		Comparisons of the transverse momentum of the top quark at 13 TeV 
		for NNLO(NLO) predictions in 5FS(4FS), presented as ratios
		to a common reference, for normalized and absolute distributions
		in (a) and (b) respectively.}
\end{figure}

\noindent \textbf{Kinematic distributions.}
Comparison of the predictions of the two schemes for various kinematic distributions of the
top quark can be enlightening, in part since there have been recommendations 
in the literature that the 4FS provides better modeling at the exclusive level~\cite{1203.6393}. 
We examine first the transverse momentum of the top quark at 13 TeV.
In Fig.~\ref{fig:pt}~(a) we show normalized cross sections at various orders with
nominal scale choices for both schemes, i.e. $\mu_{5F}=m_t/4$ and $\mu_{4F}=m_t$.
In the 5FS the LO prediction (not shown in the figure) tends to have soft spectrum 
for the transverse momentum of the top quark.
Gluon splitting at large angles can boost the top quark
in the transverse direction.
Those contributions are included at LO in the 4FS but only starting at NLO
in the 5FS. In the 5FS, we see only a modest change in shape and normalization of 
the distribution in going from NLO to NNLO.  
In Fig.~\ref{fig:pt}~(b) and~(c)  we show results for the 5FS and 
4FS respectively.  The ratio is shown of NNLO absolute cross section to the NLO predictions in 
Fig.~\ref{fig:pt}~(b) for different choices of the scale $\mu_{5F}$.  
In Fig.~\ref{fig:pt}~(c), the 
ratio is presented of the NLO and LO absolute cross sections, for various choices of $\mu_{4F}$.  
We again find that $\mu_{5F}=m_t/4$ or $m_t/2$ are the optimal choices that provide
fastest convergence in general for the transverse momentum
distribution.
Larger scales lead to enhancement of the quasi-collinear contributions
thus a softer spectrum at NLO until they are replaced by the full NNLO corrections
and vice versa.
An alternative choice could be a dynamic scale of $\mu_{5F}=H_T/4$ with
the transverse mass $H_T=(m_t^2+p_{T,top}^2)^{1/2}$.
The scale choice with transverse mass can further stabilize predictions in the
tail at high $p_T$, in qualitative agreement with the observation in
Ref.~\cite{1203.6393}, namely the collinear logarithm grows with the center of mass
energy of the hadronic system in the heavy-quark line.
Dependence of the ratios on scale choice in 4FS is seen most significantly
for the overall normalizations similar to that in Fig.~\ref{fig:totinc}.
NLO corrections in the 4FS have less impact on the shape of the distributions especially
with the choice of larger scales.
We turn next to a direct comparison of predictions of kinematic
distributions at the highest order of each scheme.
The normalized distribution on the transverse momentum of the top
quark is shown in Fig.~\ref{fig:ptrap}~(a).
We normalize the distribution to the individual total cross sections in order to 
concentrate on the shape of the distribution.  
For each distribution we plot ratios of the NNLO predictions in 5FS and NLO
predictions in 4FS to a common reference of NNLO prediction in 5FS with the
nominal scale choice $\mu_{5F}=m_t/4$.
We find remarkable agreement in shapes between the two schemes at a level of a few
percent for the kinematic region in transverse momentum considered.
The principal differences are seen close to the boundary of phase space,
e.g., at the smallest and highest transverse momenta.  
The prediction of the two schemes differ by at most
2\% for the nominal scale choices.
The spread of all predictions is within 5\% even if alternative
scale choices of $\mu_{5F}=m_t/2$ and $\mu_{4F}=m_t/2$ are chosen.

A similar comparison for the absolute distributions and for an extended
$p_T$ range is shown in Fig.~\ref{fig:ptrap}~(b).
It is interesting that the two schemes converge in the tail region of large transverse 
momentum, 
and that the normalization of the 4FS is off exactly in the region
sensitive to resummed contributions from higher orders. 
For the rapidity distribution, the spread of all predictions is at the permille level
up to a rapidity value of 2.4, and increases to at most 2\% for larger values.
This occurs because at high rapidities NNLO corrections from
the light quark line become significant and are only
included in the 5FS calculations.

\begin{figure}[h!]
	\begin{center}
		\includegraphics[width=0.45\textwidth]{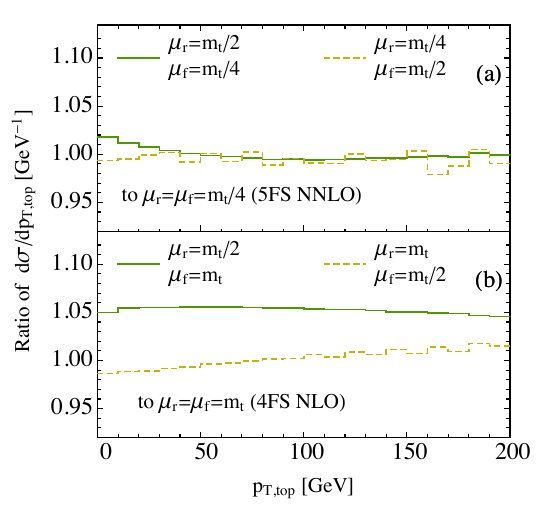}
	\end{center}
	\caption{\label{fig:ptnew}
		Scale variations of predictions for the transverse
		momentum of the top quark at 13 TeV, presented as ratios
		to the predictions with the nominal scale choice, for absolute distributions
		at NNLO(NLO) in 5FS(4FS) in (a) and (b) respectively.}
\end{figure}

We further consider the effects on the transverse momentum distribution of independent 
variations of the renormalization scale ($\mu_r$) and the factorization scale ($\mu_f$).
In Fig.~\ref{fig:ptnew}(a) we plot ratios of the predictions to those with the 
nominal scale choice when $\mu_r$ or $\mu_f$ is increased by a factor of 2,
at NNLO in 5FS.
We observe negligible changes in normalization and shape of
the distributions for variation of $\mu_r$ or $\mu_f$, 
consistent with results shown in Fig.~\ref{fig:ptrap} when the
two are varied simultaneously.
In Fig.~\ref{fig:ptnew}(b) we present similar results when 
$\mu_r$ or $\mu_f$ is decreased by a factor of 2, at NLO in 4FS.
The renormalization scale variation accounts almost fully for the
change in overall normalization of the cross sections presented earlier.
The factorization scale variations have little impact on the
total normaliztion.
They can change shape of the distribution by a few percent 
in the opposite direction to that induced by variation of renormalization
scale. 

In the conventional 5FS for single top quark production we use matrix elements with
massless bottom quarks, which is regarded as a zero-mass variable flavor number
scheme.
The power corrections from a finite bottom quark mass can be added back order by order
with the so-called general-mass variable flavor number scheme~\cite{1709.04922}.
We should not expect such power corrections to be significant for inclusive observables
in single top quark production since the top quark mass is so large~\cite{Bertone:2017djs}.
We have verified this explicitly with a NLO calculation using a simplified ACOT
scheme~\cite{Collins:1998rz,Kramer:2000hn}.
In the calculation we replace the gluon initiated matrix element in 5FS with
the LO matrix element from 4FS.
We find the finite mass corrections increase the total cross section by 0.1\%.
The impact on the shape of the transverse momentum distribution is negligible except
for a region below 20 GeV that is shifted upward by less than 1 \%.
However, the finite mass effects can be important for exclusive observables,
for instance when one measures the accompanied $b$-jet from gluon splitting,
and the bottom quark mass should be taken into account for a realistic simulation
with parton showering.
The power corrections can also be large in single top quark production at the LHeC~\cite{AbelleiraFernandez:2012cc}
where the top quarks are produced with less energy.

\begin{figure}[h!]
	\begin{center}
		\includegraphics[width=0.45\textwidth]{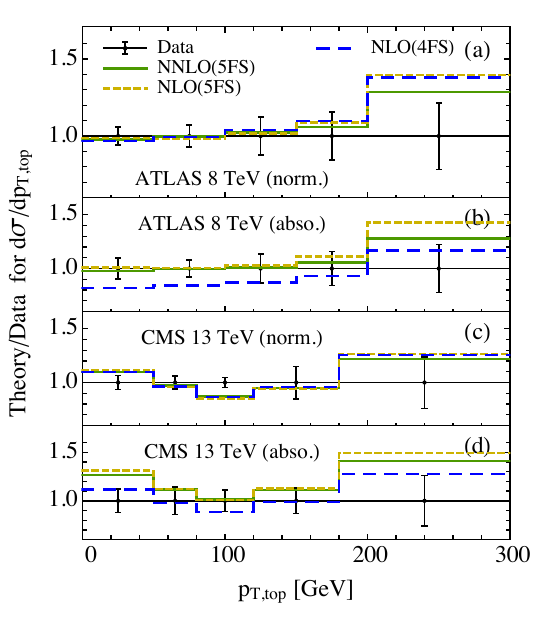}
	\end{center}
	\caption{\label{fig:ptdata}
		Comparison of predictions of absolute and normalized distributions in
		transverse momentum of the top quark to measurements from ATLAS at 8 TeV
		in (a) and (b), and to CMS at 13 TeV in (c) and (d), presented as ratios to
		central values of data. Error bars represent total experimental errors. 
		}
\end{figure}

\begin{figure}[h!]
	\begin{center}
		\includegraphics[width=0.45\textwidth]{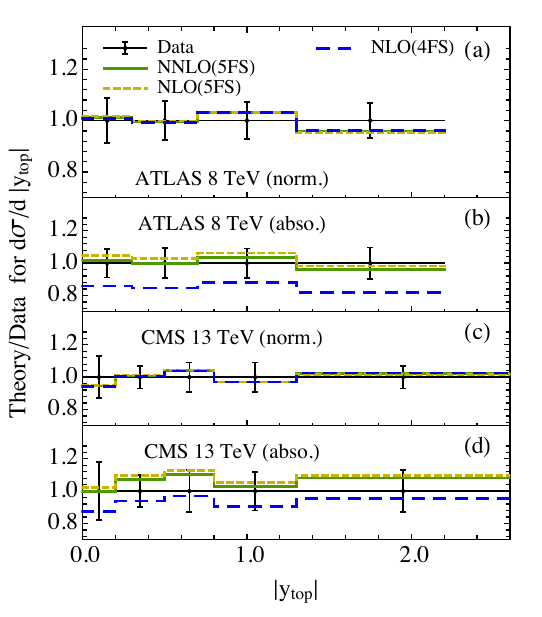}
	\end{center}
	\caption{\label{fig:rapdata}
		As in Fig.~\ref{fig:ptdata} for distributions in rapidity of the
		top quark.
		}
\end{figure}

\noindent \textbf{Comparison with data.}
The good general agreement of the theoretical predictions of the 5FS at NNLO and the 4FS 
at NLO show that uncertainties associated with scheme dependence are under control.
There are also experimental modeling uncertainties since the top quark momentum must be 
reconstructed from the kinematics of its decay products, for example, from      
semileptonic decay with an electron or muon observed in single top quark production.
These measurements are usually unfolded back to the parton level with stable
top quarks for easy comparison to theories, e.g., for a global fit of
PDFs~\cite{1912.08801,1912.09543}.
Comparison can also be made at the level of decay products if a model of top quark decay  
is included in the calculations as in Ref.~\cite{1210.2808,
1708.09405,Gao:2017goi,Liu:2018gxa,Behring:2019iiv}.
We select two measurements, one from ATLAS at 8 TeV~\cite{1702.02859} and the other from
CMS at 13 TeV~\cite{1907.08330}.
We compare predictions from both the 5FS and the 4FS with their nominal scale choices
to the measured distributions of the transverse momentum of top quark in Fig.~\ref{fig:ptdata}
and of the rapidity of the top quark in Fig.~\ref{fig:rapdata}.
In each figure we show ratios of the predictions to the central value of data for both 
absolute cross sections and normalized distributions.
For predictions of normalized distributions we normalize the bin-by-bin cross section
to the sum from all bins.
Error bars represent total experimental errors by adding statistical and systematic
errors in quadrature.


For the transverse momentum distributions shown in Fig.~\ref{fig:ptdata}~(a) and~(b), we
find very good agreement with ATLAS data for the NNLO predictions in the 5FS,
for both absolute and normalized distributions.
The NLO predictions in 4FS are systematically lower than the central values of 
ATLAS data for the absolute distribution, an aspect that can be improved if a lower
scale is used.
Regarding the case of CMS, we find none of the theoretical curves describes the CMS data 
particularly well for the normalized distribution in Fig.~\ref{fig:ptdata}~(c).
The differences of the predictions in 5FS and 4FS are much
smaller than the experimental errors for the normalized distribution.
Interestingly the CMS data on the absolute distribution in Fig.~\ref{fig:ptdata}~(d) seem 
to agree better with the NLO prediction in the 4FS for the
overall normalization.
\footnote{A decay branching to two lepton families of top quark is applied in order
to compare with the CMS measurement on top-quark distributions at parton level.} 
This is opposite to the case of the total inclusive
cross sections at 13 TeV~\cite{Aaboud:2016ymp,Sirunyan:2016cdg,Sirunyan:2018rlu} 
which agree better with predictions in the 5FS.

For rapidity distributions shown in Fig.~\ref{fig:rapdata}, all
predictions agree quite well with the data on normalized distributions.
The normalization of predictions in the 4FS are again lower than the ATLAS central data.
We find the overall normalization of CMS data on the rapidity distribution
is larger by 6\% compared to data on the transverse momentum distribution.
Comparing Fig.~\ref{fig:ptdata}~(d)
and Fig.~\ref{fig:rapdata}~(d) we see that the predictions from 4FS are higher than
the central values of data on average for transverse momentum
and much lower than data for rapidity.

We are left puzzled by what may be inconsistencies within the CMS data set and refrain 
from drawing stronger conclusions.

\noindent \textbf{Summary.}
We study the modeling of $t$-channel single top quark production at the LHC
at the highest perturbative order available in both a 5-flavor
and 4-flavor scheme.
We find excellent agreement between the two schemes for predictions
of the shape of kinematic distributions of the top quark.
The 5FS further exhibits strong stability of predictions of  the 
normalization and distributions, and are superior to predictions from 4FS when
evaluated at comparable orders in perturbation theory.
Our comparisons with current data on
top quark distributions show good agreement with ATLAS measurements
but some discrepancies with CMS.
The perturbative uncertainty reaches a few percent for both inclusive
cross sections and distributions with NNLO predictions in 5FS.
Our results point the way toward the precision study of single top quark
production in future studies with LHC data.

\begin{acknowledgments}
Work in the High Energy Physics Division at Argonne is supported in part by the U.S. Department of Energy under 
Contract No. DE-AC02-06CH11357.
The work of J.~Gao was sponsored by the National Natural
Science Foundation of China under the Grant No. 11875189 and No.11835005.
The authors thank Hua Xing Zhu
for useful discussions and for collaboration in the early stages of
this work.
\\
\end{acknowledgments}
\bibliographystyle{apsrev}
\bibliography{hfstop}

\begin{thebibliography}{67}
\expandafter\ifx\csname natexlab\endcsname\relax\def\natexlab#1{#1}\fi
\expandafter\ifx\csname bibnamefont\endcsname\relax
  \def\bibnamefont#1{#1}\fi
\expandafter\ifx\csname bibfnamefont\endcsname\relax
  \def\bibfnamefont#1{#1}\fi
\expandafter\ifx\csname citenamefont\endcsname\relax
  \def\citenamefont#1{#1}\fi
\expandafter\ifx\csname url\endcsname\relax
  \def\url#1{\texttt{#1}}\fi
\expandafter\ifx\csname urlprefix\endcsname\relax\def\urlprefix{URL }\fi
\providecommand{\bibinfo}[2]{#2}
\providecommand{\eprint}[2][]{\url{#2}}

\bibitem[{\citenamefont{Alekhin
  et~al.}(2016{\natexlab{a}})\citenamefont{Alekhin, Moch, and
  Thier}}]{Alekhin:2016jjz}
\bibinfo{author}{\bibfnamefont{S.}~\bibnamefont{Alekhin}},
  \bibinfo{author}{\bibfnamefont{S.}~\bibnamefont{Moch}}, \bibnamefont{and}
  \bibinfo{author}{\bibfnamefont{S.}~\bibnamefont{Thier}},
  \bibinfo{journal}{Phys. Lett.} \textbf{\bibinfo{volume}{B763}},
  \bibinfo{pages}{341} (\bibinfo{year}{2016}{\natexlab{a}}),
  \eprint{1608.05212}.

\bibitem[{\citenamefont{Sirunyan
  et~al.}(2017{\natexlab{a}})}]{Sirunyan:2017huu}
\bibinfo{author}{\bibfnamefont{A.~M.} \bibnamefont{Sirunyan}}
  \bibnamefont{et~al.} (\bibinfo{collaboration}{CMS}), \bibinfo{journal}{Eur.
  Phys. J.} \textbf{\bibinfo{volume}{C77}}, \bibinfo{pages}{354}
  (\bibinfo{year}{2017}{\natexlab{a}}), \eprint{1703.02530}.

\bibitem[{\citenamefont{Brucherseifer et~al.}(2014)\citenamefont{Brucherseifer,
  Caola, and Melnikov}}]{1404.7116}
\bibinfo{author}{\bibfnamefont{M.}~\bibnamefont{Brucherseifer}},
  \bibinfo{author}{\bibfnamefont{F.}~\bibnamefont{Caola}}, \bibnamefont{and}
  \bibinfo{author}{\bibfnamefont{K.}~\bibnamefont{Melnikov}},
  \bibinfo{journal}{Phys. Lett.} \textbf{\bibinfo{volume}{B736}},
  \bibinfo{pages}{58} (\bibinfo{year}{2014}), \eprint{1404.7116}.

\bibitem[{\citenamefont{Alekhin
  et~al.}(2016{\natexlab{b}})\citenamefont{Alekhin, Blümlein, Moch, and
  Plačakytė}}]{Alekhin:2015cza}
\bibinfo{author}{\bibfnamefont{S.}~\bibnamefont{Alekhin}},
  \bibinfo{author}{\bibfnamefont{J.}~\bibnamefont{Blümlein}},
  \bibinfo{author}{\bibfnamefont{S.}~\bibnamefont{Moch}}, \bibnamefont{and}
  \bibinfo{author}{\bibfnamefont{R.}~\bibnamefont{Plačakytė}},
  \bibinfo{journal}{Phys. Rev.} \textbf{\bibinfo{volume}{D94}},
  \bibinfo{pages}{114038} (\bibinfo{year}{2016}{\natexlab{b}}),
  \eprint{1508.07923}.

\bibitem[{\citenamefont{Berger et~al.}(2016{\natexlab{a}})\citenamefont{Berger,
  Gao, Yuan, and Zhu}}]{Berger:2016oht}
\bibinfo{author}{\bibfnamefont{E.~L.} \bibnamefont{Berger}},
  \bibinfo{author}{\bibfnamefont{J.}~\bibnamefont{Gao}},
  \bibinfo{author}{\bibfnamefont{C.~P.} \bibnamefont{Yuan}}, \bibnamefont{and}
  \bibinfo{author}{\bibfnamefont{H.~X.} \bibnamefont{Zhu}},
  \bibinfo{journal}{Phys. Rev.} \textbf{\bibinfo{volume}{D94}},
  \bibinfo{pages}{071501} (\bibinfo{year}{2016}{\natexlab{a}}),
  \eprint{1606.08463}.

\bibitem[{\citenamefont{Nocera et~al.}(2019)\citenamefont{Nocera, Ubiali, and
  Voisey}}]{1912.09543}
\bibinfo{author}{\bibfnamefont{E.~R.} \bibnamefont{Nocera}},
  \bibinfo{author}{\bibfnamefont{M.}~\bibnamefont{Ubiali}}, \bibnamefont{and}
  \bibinfo{author}{\bibfnamefont{C.}~\bibnamefont{Voisey}}
  (\bibinfo{year}{2019}), \eprint{1912.09543}.

\bibitem[{\citenamefont{He and Yuan}(1999)}]{He:1998ie}
\bibinfo{author}{\bibfnamefont{H.-J.} \bibnamefont{He}} \bibnamefont{and}
  \bibinfo{author}{\bibfnamefont{C.-P.} \bibnamefont{Yuan}},
  \bibinfo{journal}{Phys. Rev. Lett.} \textbf{\bibinfo{volume}{83}},
  \bibinfo{pages}{28} (\bibinfo{year}{1999}), \eprint{hep-ph/9810367}.

\bibitem[{\citenamefont{Tait and Yuan}(2000)}]{hep-ph/0007298}
\bibinfo{author}{\bibfnamefont{T.~M.~P.} \bibnamefont{Tait}} \bibnamefont{and}
  \bibinfo{author}{\bibfnamefont{C.~P.} \bibnamefont{Yuan}},
  \bibinfo{journal}{Phys. Rev.} \textbf{\bibinfo{volume}{D63}},
  \bibinfo{pages}{014018} (\bibinfo{year}{2000}), \eprint{hep-ph/0007298}.

\bibitem[{\citenamefont{Campbell et~al.}(2009)\citenamefont{Campbell, Frederix,
  Maltoni, and Tramontano}}]{0903.0005}
\bibinfo{author}{\bibfnamefont{J.~M.} \bibnamefont{Campbell}},
  \bibinfo{author}{\bibfnamefont{R.}~\bibnamefont{Frederix}},
  \bibinfo{author}{\bibfnamefont{F.}~\bibnamefont{Maltoni}}, \bibnamefont{and}
  \bibinfo{author}{\bibfnamefont{F.}~\bibnamefont{Tramontano}},
  \bibinfo{journal}{Phys. Rev. Lett.} \textbf{\bibinfo{volume}{102}},
  \bibinfo{pages}{182003} (\bibinfo{year}{2009}), \eprint{0903.0005}.

\bibitem[{\citenamefont{Maltoni et~al.}(2012)\citenamefont{Maltoni, Ridolfi,
  and Ubiali}}]{1203.6393}
\bibinfo{author}{\bibfnamefont{F.}~\bibnamefont{Maltoni}},
  \bibinfo{author}{\bibfnamefont{G.}~\bibnamefont{Ridolfi}}, \bibnamefont{and}
  \bibinfo{author}{\bibfnamefont{M.}~\bibnamefont{Ubiali}},
  \bibinfo{journal}{JHEP} \textbf{\bibinfo{volume}{07}}, \bibinfo{pages}{022}
  (\bibinfo{year}{2012}), \bibinfo{note}{[Erratum: JHEP04,095(2013)]},
  \eprint{1203.6393}.

\bibitem[{\citenamefont{Bothmann et~al.}(2018)\citenamefont{Bothmann, Krauss,
  and Schönherr}}]{1711.02568}
\bibinfo{author}{\bibfnamefont{E.}~\bibnamefont{Bothmann}},
  \bibinfo{author}{\bibfnamefont{F.}~\bibnamefont{Krauss}}, \bibnamefont{and}
  \bibinfo{author}{\bibfnamefont{M.}~\bibnamefont{Schönherr}},
  \bibinfo{journal}{Eur. Phys. J.} \textbf{\bibinfo{volume}{C78}},
  \bibinfo{pages}{220} (\bibinfo{year}{2018}), \eprint{1711.02568}.

\bibitem[{\citenamefont{Aaboud et~al.}(2017{\natexlab{a}})}]{1702.02859}
\bibinfo{author}{\bibfnamefont{M.}~\bibnamefont{Aaboud}} \bibnamefont{et~al.}
  (\bibinfo{collaboration}{ATLAS}), \bibinfo{journal}{Eur. Phys. J.}
  \textbf{\bibinfo{volume}{C77}}, \bibinfo{pages}{531}
  (\bibinfo{year}{2017}{\natexlab{a}}), \eprint{1702.02859}.

\bibitem[{\citenamefont{Sirunyan
  et~al.}(2020{\natexlab{a}})}]{Sirunyan:2018rlu}
\bibinfo{author}{\bibfnamefont{A.~M.} \bibnamefont{Sirunyan}}
  \bibnamefont{et~al.} (\bibinfo{collaboration}{CMS}), \bibinfo{journal}{Phys.
  Lett.} \textbf{\bibinfo{volume}{B800}}, \bibinfo{pages}{135042}
  (\bibinfo{year}{2020}{\natexlab{a}}), \eprint{1812.10514}.

\bibitem[{\citenamefont{Aaboud et~al.}(2019)}]{Aaboud:2019pkc}
\bibinfo{author}{\bibfnamefont{M.}~\bibnamefont{Aaboud}} \bibnamefont{et~al.}
  (\bibinfo{collaboration}{ATLAS, CMS}), \bibinfo{journal}{JHEP}
  \textbf{\bibinfo{volume}{05}}, \bibinfo{pages}{088} (\bibinfo{year}{2019}),
  \eprint{1902.07158}.

\bibitem[{\citenamefont{Bordes and van Eijk}(1995)}]{NUPHA.B435.23}
\bibinfo{author}{\bibfnamefont{G.}~\bibnamefont{Bordes}} \bibnamefont{and}
  \bibinfo{author}{\bibfnamefont{B.}~\bibnamefont{van Eijk}},
  \bibinfo{journal}{Nucl. Phys.} \textbf{\bibinfo{volume}{B435}},
  \bibinfo{pages}{23} (\bibinfo{year}{1995}).

\bibitem[{\citenamefont{Pittau}(1996)}]{hep-ph/9603265}
\bibinfo{author}{\bibfnamefont{R.}~\bibnamefont{Pittau}},
  \bibinfo{journal}{Phys. Lett.} \textbf{\bibinfo{volume}{B386}},
  \bibinfo{pages}{397} (\bibinfo{year}{1996}), \eprint{hep-ph/9603265}.

\bibitem[{\citenamefont{Stelzer et~al.}(1997)\citenamefont{Stelzer, Sullivan,
  and Willenbrock}}]{hep-ph/9705398}
\bibinfo{author}{\bibfnamefont{T.}~\bibnamefont{Stelzer}},
  \bibinfo{author}{\bibfnamefont{Z.}~\bibnamefont{Sullivan}}, \bibnamefont{and}
  \bibinfo{author}{\bibfnamefont{S.}~\bibnamefont{Willenbrock}},
  \bibinfo{journal}{Phys. Rev.} \textbf{\bibinfo{volume}{D56}},
  \bibinfo{pages}{5919} (\bibinfo{year}{1997}), \eprint{hep-ph/9705398}.

\bibitem[{\citenamefont{Stelzer et~al.}(1998)\citenamefont{Stelzer, Sullivan,
  and Willenbrock}}]{hep-ph/9807340}
\bibinfo{author}{\bibfnamefont{T.}~\bibnamefont{Stelzer}},
  \bibinfo{author}{\bibfnamefont{Z.}~\bibnamefont{Sullivan}}, \bibnamefont{and}
  \bibinfo{author}{\bibfnamefont{S.}~\bibnamefont{Willenbrock}},
  \bibinfo{journal}{Phys. Rev.} \textbf{\bibinfo{volume}{D58}},
  \bibinfo{pages}{094021} (\bibinfo{year}{1998}), \eprint{hep-ph/9807340}.

\bibitem[{\citenamefont{Harris et~al.}(2001)\citenamefont{Harris, Laenen, Phaf,
  Sullivan, and Weinzierl}}]{hep-ph/0102126}
\bibinfo{author}{\bibfnamefont{B.~W.} \bibnamefont{Harris}},
  \bibinfo{author}{\bibfnamefont{E.}~\bibnamefont{Laenen}},
  \bibinfo{author}{\bibfnamefont{L.}~\bibnamefont{Phaf}},
  \bibinfo{author}{\bibfnamefont{Z.}~\bibnamefont{Sullivan}}, \bibnamefont{and}
  \bibinfo{author}{\bibfnamefont{S.}~\bibnamefont{Weinzierl}},
  \bibinfo{journal}{Int. J. Mod. Phys.} \textbf{\bibinfo{volume}{A16S1A}},
  \bibinfo{pages}{379} (\bibinfo{year}{2001}), \eprint{hep-ph/0102126}.

\bibitem[{\citenamefont{Harris et~al.}(2002)\citenamefont{Harris, Laenen, Phaf,
  Sullivan, and Weinzierl}}]{hep-ph/0207055}
\bibinfo{author}{\bibfnamefont{B.~W.} \bibnamefont{Harris}},
  \bibinfo{author}{\bibfnamefont{E.}~\bibnamefont{Laenen}},
  \bibinfo{author}{\bibfnamefont{L.}~\bibnamefont{Phaf}},
  \bibinfo{author}{\bibfnamefont{Z.}~\bibnamefont{Sullivan}}, \bibnamefont{and}
  \bibinfo{author}{\bibfnamefont{S.}~\bibnamefont{Weinzierl}},
  \bibinfo{journal}{Phys. Rev.} \textbf{\bibinfo{volume}{D66}},
  \bibinfo{pages}{054024} (\bibinfo{year}{2002}), \eprint{hep-ph/0207055}.

\bibitem[{\citenamefont{Sullivan}(2004)}]{Sullivan:2004ie}
\bibinfo{author}{\bibfnamefont{Z.}~\bibnamefont{Sullivan}},
  \bibinfo{journal}{Phys. Rev.} \textbf{\bibinfo{volume}{D70}},
  \bibinfo{pages}{114012} (\bibinfo{year}{2004}), \eprint{hep-ph/0408049}.

\bibitem[{\citenamefont{Campbell et~al.}(2004)\citenamefont{Campbell, Ellis,
  and Tramontano}}]{hep-ph/0408158}
\bibinfo{author}{\bibfnamefont{J.~M.} \bibnamefont{Campbell}},
  \bibinfo{author}{\bibfnamefont{R.~K.} \bibnamefont{Ellis}}, \bibnamefont{and}
  \bibinfo{author}{\bibfnamefont{F.}~\bibnamefont{Tramontano}},
  \bibinfo{journal}{Phys. Rev.} \textbf{\bibinfo{volume}{D70}},
  \bibinfo{pages}{094012} (\bibinfo{year}{2004}), \eprint{hep-ph/0408158}.

\bibitem[{\citenamefont{Sullivan}(2005)}]{hep-ph/0510224}
\bibinfo{author}{\bibfnamefont{Z.}~\bibnamefont{Sullivan}},
  \bibinfo{journal}{Phys. Rev.} \textbf{\bibinfo{volume}{D72}},
  \bibinfo{pages}{094034} (\bibinfo{year}{2005}), \eprint{hep-ph/0510224}.

\bibitem[{\citenamefont{Cao et~al.}(2005)\citenamefont{Cao, Schwienhorst,
  Benitez, Brock, and Yuan}}]{hep-ph/0504230}
\bibinfo{author}{\bibfnamefont{Q.-H.} \bibnamefont{Cao}},
  \bibinfo{author}{\bibfnamefont{R.}~\bibnamefont{Schwienhorst}},
  \bibinfo{author}{\bibfnamefont{J.~A.} \bibnamefont{Benitez}},
  \bibinfo{author}{\bibfnamefont{R.}~\bibnamefont{Brock}}, \bibnamefont{and}
  \bibinfo{author}{\bibfnamefont{C.~P.} \bibnamefont{Yuan}},
  \bibinfo{journal}{Phys. Rev.} \textbf{\bibinfo{volume}{D72}},
  \bibinfo{pages}{094027} (\bibinfo{year}{2005}), \eprint{hep-ph/0504230}.

\bibitem[{\citenamefont{Falgari et~al.}(2010)\citenamefont{Falgari, Mellor, and
  Signer}}]{1007.0893}
\bibinfo{author}{\bibfnamefont{P.}~\bibnamefont{Falgari}},
  \bibinfo{author}{\bibfnamefont{P.}~\bibnamefont{Mellor}}, \bibnamefont{and}
  \bibinfo{author}{\bibfnamefont{A.}~\bibnamefont{Signer}},
  \bibinfo{journal}{Phys. Rev.} \textbf{\bibinfo{volume}{D82}},
  \bibinfo{pages}{054028} (\bibinfo{year}{2010}), \eprint{1007.0893}.

\bibitem[{\citenamefont{Schwienhorst et~al.}(2011)\citenamefont{Schwienhorst,
  Yuan, Mueller, and Cao}}]{1012.5132}
\bibinfo{author}{\bibfnamefont{R.}~\bibnamefont{Schwienhorst}},
  \bibinfo{author}{\bibfnamefont{C.~P.} \bibnamefont{Yuan}},
  \bibinfo{author}{\bibfnamefont{C.}~\bibnamefont{Mueller}}, \bibnamefont{and}
  \bibinfo{author}{\bibfnamefont{Q.-H.} \bibnamefont{Cao}},
  \bibinfo{journal}{Phys. Rev.} \textbf{\bibinfo{volume}{D83}},
  \bibinfo{pages}{034019} (\bibinfo{year}{2011}), \eprint{1012.5132}.

\bibitem[{\citenamefont{Falgari et~al.}(2011)\citenamefont{Falgari, Giannuzzi,
  Mellor, and Signer}}]{1102.5267}
\bibinfo{author}{\bibfnamefont{P.}~\bibnamefont{Falgari}},
  \bibinfo{author}{\bibfnamefont{F.}~\bibnamefont{Giannuzzi}},
  \bibinfo{author}{\bibfnamefont{P.}~\bibnamefont{Mellor}}, \bibnamefont{and}
  \bibinfo{author}{\bibfnamefont{A.}~\bibnamefont{Signer}},
  \bibinfo{journal}{Phys. Rev.} \textbf{\bibinfo{volume}{D83}},
  \bibinfo{pages}{094013} (\bibinfo{year}{2011}), \eprint{1102.5267}.

\bibitem[{\citenamefont{Papanastasiou et~al.}(2013)\citenamefont{Papanastasiou,
  Frederix, Frixione, Hirschi, and Maltoni}}]{1305.7088}
\bibinfo{author}{\bibfnamefont{A.~S.} \bibnamefont{Papanastasiou}},
  \bibinfo{author}{\bibfnamefont{R.}~\bibnamefont{Frederix}},
  \bibinfo{author}{\bibfnamefont{S.}~\bibnamefont{Frixione}},
  \bibinfo{author}{\bibfnamefont{V.}~\bibnamefont{Hirschi}}, \bibnamefont{and}
  \bibinfo{author}{\bibfnamefont{F.}~\bibnamefont{Maltoni}},
  \bibinfo{journal}{Phys. Lett.} \textbf{\bibinfo{volume}{B726}},
  \bibinfo{pages}{223} (\bibinfo{year}{2013}), \eprint{1305.7088}.

\bibitem[{\citenamefont{Kant et~al.}(2015)\citenamefont{Kant, Kind, Kintscher,
  Lohse, Martini, Mölbitz, Rieck, and Uwer}}]{1406.4403}
\bibinfo{author}{\bibfnamefont{P.}~\bibnamefont{Kant}},
  \bibinfo{author}{\bibfnamefont{O.~M.} \bibnamefont{Kind}},
  \bibinfo{author}{\bibfnamefont{T.}~\bibnamefont{Kintscher}},
  \bibinfo{author}{\bibfnamefont{T.}~\bibnamefont{Lohse}},
  \bibinfo{author}{\bibfnamefont{T.}~\bibnamefont{Martini}},
  \bibinfo{author}{\bibfnamefont{S.}~\bibnamefont{Mölbitz}},
  \bibinfo{author}{\bibfnamefont{P.}~\bibnamefont{Rieck}}, \bibnamefont{and}
  \bibinfo{author}{\bibfnamefont{P.}~\bibnamefont{Uwer}},
  \bibinfo{journal}{Comput. Phys. Commun.} \textbf{\bibinfo{volume}{191}},
  \bibinfo{pages}{74} (\bibinfo{year}{2015}), \eprint{1406.4403}.

\bibitem[{\citenamefont{Carrazza et~al.}(2018)\citenamefont{Carrazza, Frederix,
  Hamilton, and Zanderighi}}]{Carrazza:2018mix}
\bibinfo{author}{\bibfnamefont{S.}~\bibnamefont{Carrazza}},
  \bibinfo{author}{\bibfnamefont{R.}~\bibnamefont{Frederix}},
  \bibinfo{author}{\bibfnamefont{K.}~\bibnamefont{Hamilton}}, \bibnamefont{and}
  \bibinfo{author}{\bibfnamefont{G.}~\bibnamefont{Zanderighi}},
  \bibinfo{journal}{JHEP} \textbf{\bibinfo{volume}{09}}, \bibinfo{pages}{108}
  (\bibinfo{year}{2018}), \eprint{1805.09855}.

\bibitem[{\citenamefont{Berger et~al.}(2017)\citenamefont{Berger, Gao, and
  Zhu}}]{1708.09405}
\bibinfo{author}{\bibfnamefont{E.~L.} \bibnamefont{Berger}},
  \bibinfo{author}{\bibfnamefont{J.}~\bibnamefont{Gao}}, \bibnamefont{and}
  \bibinfo{author}{\bibfnamefont{H.~X.} \bibnamefont{Zhu}},
  \bibinfo{journal}{JHEP} \textbf{\bibinfo{volume}{11}}, \bibinfo{pages}{158}
  (\bibinfo{year}{2017}), \eprint{1708.09405}.

\bibitem[{\citenamefont{Frederix et~al.}(2019)\citenamefont{Frederix, Pagani,
  and Tsinikos}}]{1907.12586}
\bibinfo{author}{\bibfnamefont{R.}~\bibnamefont{Frederix}},
  \bibinfo{author}{\bibfnamefont{D.}~\bibnamefont{Pagani}}, \bibnamefont{and}
  \bibinfo{author}{\bibfnamefont{I.}~\bibnamefont{Tsinikos}},
  \bibinfo{journal}{JHEP} \textbf{\bibinfo{volume}{09}}, \bibinfo{pages}{122}
  (\bibinfo{year}{2019}), \eprint{1907.12586}.

\bibitem[{\citenamefont{Wang et~al.}(2010)\citenamefont{Wang, Li, Zhu, and
  Zhang}}]{1010.4509}
\bibinfo{author}{\bibfnamefont{J.}~\bibnamefont{Wang}},
  \bibinfo{author}{\bibfnamefont{C.~S.} \bibnamefont{Li}},
  \bibinfo{author}{\bibfnamefont{H.~X.} \bibnamefont{Zhu}}, \bibnamefont{and}
  \bibinfo{author}{\bibfnamefont{J.~J.} \bibnamefont{Zhang}}
  (\bibinfo{year}{2010}), \eprint{1010.4509}.

\bibitem[{\citenamefont{Kidonakis}(2011)}]{1103.2792}
\bibinfo{author}{\bibfnamefont{N.}~\bibnamefont{Kidonakis}},
  \bibinfo{journal}{Phys. Rev.} \textbf{\bibinfo{volume}{D83}},
  \bibinfo{pages}{091503} (\bibinfo{year}{2011}), \eprint{1103.2792}.

\bibitem[{\citenamefont{Wang et~al.}(2013)\citenamefont{Wang, Li, and
  Zhu}}]{1210.7698}
\bibinfo{author}{\bibfnamefont{J.}~\bibnamefont{Wang}},
  \bibinfo{author}{\bibfnamefont{C.~S.} \bibnamefont{Li}}, \bibnamefont{and}
  \bibinfo{author}{\bibfnamefont{H.~X.} \bibnamefont{Zhu}},
  \bibinfo{journal}{Phys. Rev.} \textbf{\bibinfo{volume}{D87}},
  \bibinfo{pages}{034030} (\bibinfo{year}{2013}), \eprint{1210.7698}.

\bibitem[{\citenamefont{Kidonakis}(2016)}]{1510.06361}
\bibinfo{author}{\bibfnamefont{N.}~\bibnamefont{Kidonakis}},
  \bibinfo{journal}{Phys. Rev.} \textbf{\bibinfo{volume}{D93}},
  \bibinfo{pages}{054022} (\bibinfo{year}{2016}), \eprint{1510.06361}.

\bibitem[{\citenamefont{Cao et~al.}(2018)\citenamefont{Cao, Sun, Yan, Yuan, and
  Yuan}}]{1801.09656}
\bibinfo{author}{\bibfnamefont{Q.-H.} \bibnamefont{Cao}},
  \bibinfo{author}{\bibfnamefont{P.}~\bibnamefont{Sun}},
  \bibinfo{author}{\bibfnamefont{B.}~\bibnamefont{Yan}},
  \bibinfo{author}{\bibfnamefont{C.~P.} \bibnamefont{Yuan}}, \bibnamefont{and}
  \bibinfo{author}{\bibfnamefont{F.}~\bibnamefont{Yuan}},
  \bibinfo{journal}{Phys. Rev.} \textbf{\bibinfo{volume}{D98}},
  \bibinfo{pages}{054032} (\bibinfo{year}{2018}), \eprint{1801.09656}.

\bibitem[{\citenamefont{Kidonakis}(2019)}]{Kidonakis:2019nqa}
\bibinfo{author}{\bibfnamefont{N.}~\bibnamefont{Kidonakis}},
  \bibinfo{journal}{Phys. Rev.} \textbf{\bibinfo{volume}{D99}},
  \bibinfo{pages}{074024} (\bibinfo{year}{2019}), \eprint{1901.09928}.

\bibitem[{\citenamefont{Cao et~al.}(2019)\citenamefont{Cao, Sun, Yan, Yuan, and
  Yuan}}]{Cao:2019uor}
\bibinfo{author}{\bibfnamefont{Q.-H.} \bibnamefont{Cao}},
  \bibinfo{author}{\bibfnamefont{P.}~\bibnamefont{Sun}},
  \bibinfo{author}{\bibfnamefont{B.}~\bibnamefont{Yan}},
  \bibinfo{author}{\bibfnamefont{C.~P.} \bibnamefont{Yuan}}, \bibnamefont{and}
  \bibinfo{author}{\bibfnamefont{F.}~\bibnamefont{Yuan}}
  (\bibinfo{year}{2019}), \eprint{1902.09336}.

\bibitem[{\citenamefont{Frixione et~al.}(2006)\citenamefont{Frixione, Laenen,
  Motylinski, and Webber}}]{hep-ph/0512250}
\bibinfo{author}{\bibfnamefont{S.}~\bibnamefont{Frixione}},
  \bibinfo{author}{\bibfnamefont{E.}~\bibnamefont{Laenen}},
  \bibinfo{author}{\bibfnamefont{P.}~\bibnamefont{Motylinski}},
  \bibnamefont{and} \bibinfo{author}{\bibfnamefont{B.~R.}
  \bibnamefont{Webber}}, \bibinfo{journal}{JHEP} \textbf{\bibinfo{volume}{03}},
  \bibinfo{pages}{092} (\bibinfo{year}{2006}), \eprint{hep-ph/0512250}.

\bibitem[{\citenamefont{Alioli et~al.}(2009)\citenamefont{Alioli, Nason,
  Oleari, and Re}}]{0907.4076}
\bibinfo{author}{\bibfnamefont{S.}~\bibnamefont{Alioli}},
  \bibinfo{author}{\bibfnamefont{P.}~\bibnamefont{Nason}},
  \bibinfo{author}{\bibfnamefont{C.}~\bibnamefont{Oleari}}, \bibnamefont{and}
  \bibinfo{author}{\bibfnamefont{E.}~\bibnamefont{Re}}, \bibinfo{journal}{JHEP}
  \textbf{\bibinfo{volume}{09}}, \bibinfo{pages}{111} (\bibinfo{year}{2009}),
  \bibinfo{note}{[Erratum: JHEP02,011(2010)]}, \eprint{0907.4076}.

\bibitem[{\citenamefont{Frederix et~al.}(2012)\citenamefont{Frederix, Re, and
  Torrielli}}]{1207.5391}
\bibinfo{author}{\bibfnamefont{R.}~\bibnamefont{Frederix}},
  \bibinfo{author}{\bibfnamefont{E.}~\bibnamefont{Re}}, \bibnamefont{and}
  \bibinfo{author}{\bibfnamefont{P.}~\bibnamefont{Torrielli}},
  \bibinfo{journal}{JHEP} \textbf{\bibinfo{volume}{09}}, \bibinfo{pages}{130}
  (\bibinfo{year}{2012}), \eprint{1207.5391}.

\bibitem[{\citenamefont{Frederix et~al.}(2016)\citenamefont{Frederix, Frixione,
  Papanastasiou, Prestel, and Torrielli}}]{1603.01178}
\bibinfo{author}{\bibfnamefont{R.}~\bibnamefont{Frederix}},
  \bibinfo{author}{\bibfnamefont{S.}~\bibnamefont{Frixione}},
  \bibinfo{author}{\bibfnamefont{A.~S.} \bibnamefont{Papanastasiou}},
  \bibinfo{author}{\bibfnamefont{S.}~\bibnamefont{Prestel}}, \bibnamefont{and}
  \bibinfo{author}{\bibfnamefont{P.}~\bibnamefont{Torrielli}},
  \bibinfo{journal}{JHEP} \textbf{\bibinfo{volume}{06}}, \bibinfo{pages}{027}
  (\bibinfo{year}{2016}), \eprint{1603.01178}.

\bibitem[{\citenamefont{Stewart et~al.}(2010)\citenamefont{Stewart, Tackmann,
  and Waalewijn}}]{Stewart:2010tn}
\bibinfo{author}{\bibfnamefont{I.~W.} \bibnamefont{Stewart}},
  \bibinfo{author}{\bibfnamefont{F.~J.} \bibnamefont{Tackmann}},
  \bibnamefont{and} \bibinfo{author}{\bibfnamefont{W.~J.}
  \bibnamefont{Waalewijn}}, \bibinfo{journal}{Phys. Rev. Lett.}
  \textbf{\bibinfo{volume}{105}}, \bibinfo{pages}{092002}
  (\bibinfo{year}{2010}), \eprint{1004.2489}.

\bibitem[{\citenamefont{Boughezal et~al.}(2015)\citenamefont{Boughezal, Focke,
  Liu, and Petriello}}]{Boughezal:2015dva}
\bibinfo{author}{\bibfnamefont{R.}~\bibnamefont{Boughezal}},
  \bibinfo{author}{\bibfnamefont{C.}~\bibnamefont{Focke}},
  \bibinfo{author}{\bibfnamefont{X.}~\bibnamefont{Liu}}, \bibnamefont{and}
  \bibinfo{author}{\bibfnamefont{F.}~\bibnamefont{Petriello}},
  \bibinfo{journal}{Phys. Rev. Lett.} \textbf{\bibinfo{volume}{115}},
  \bibinfo{pages}{062002} (\bibinfo{year}{2015}), \eprint{1504.02131}.

\bibitem[{\citenamefont{Gaunt et~al.}(2015)\citenamefont{Gaunt, Stahlhofen,
  Tackmann, and Walsh}}]{Gaunt:2015pea}
\bibinfo{author}{\bibfnamefont{J.}~\bibnamefont{Gaunt}},
  \bibinfo{author}{\bibfnamefont{M.}~\bibnamefont{Stahlhofen}},
  \bibinfo{author}{\bibfnamefont{F.~J.} \bibnamefont{Tackmann}},
  \bibnamefont{and} \bibinfo{author}{\bibfnamefont{J.~R.} \bibnamefont{Walsh}},
  \bibinfo{journal}{JHEP} \textbf{\bibinfo{volume}{09}}, \bibinfo{pages}{058}
  (\bibinfo{year}{2015}), \eprint{1505.04794}.

\bibitem[{\citenamefont{Berger et~al.}(2016{\natexlab{b}})\citenamefont{Berger,
  Gao, Li, Liu, and Zhu}}]{Berger:2016inr}
\bibinfo{author}{\bibfnamefont{E.~L.} \bibnamefont{Berger}},
  \bibinfo{author}{\bibfnamefont{J.}~\bibnamefont{Gao}},
  \bibinfo{author}{\bibfnamefont{C.~S.} \bibnamefont{Li}},
  \bibinfo{author}{\bibfnamefont{Z.~L.} \bibnamefont{Liu}}, \bibnamefont{and}
  \bibinfo{author}{\bibfnamefont{H.~X.} \bibnamefont{Zhu}},
  \bibinfo{journal}{Phys. Rev. Lett.} \textbf{\bibinfo{volume}{116}},
  \bibinfo{pages}{212002} (\bibinfo{year}{2016}{\natexlab{b}}),
  \eprint{1601.05430}.

\bibitem[{\citenamefont{Cacciari et~al.}(2015)\citenamefont{Cacciari, Dreyer,
  Karlberg, Salam, and Zanderighi}}]{1506.02660}
\bibinfo{author}{\bibfnamefont{M.}~\bibnamefont{Cacciari}},
  \bibinfo{author}{\bibfnamefont{F.~A.} \bibnamefont{Dreyer}},
  \bibinfo{author}{\bibfnamefont{A.}~\bibnamefont{Karlberg}},
  \bibinfo{author}{\bibfnamefont{G.~P.} \bibnamefont{Salam}}, \bibnamefont{and}
  \bibinfo{author}{\bibfnamefont{G.}~\bibnamefont{Zanderighi}},
  \bibinfo{journal}{Phys. Rev. Lett.} \textbf{\bibinfo{volume}{115}},
  \bibinfo{pages}{082002} (\bibinfo{year}{2015}), \bibinfo{note}{[Erratum:
  Phys. Rev. Lett.120,no.13,139901(2018)]}, \eprint{1506.02660}.

\bibitem[{\citenamefont{Campbell et~al.}(2016)\citenamefont{Campbell, Ellis,
  and Williams}}]{Campbell:2016jau}
\bibinfo{author}{\bibfnamefont{J.~M.} \bibnamefont{Campbell}},
  \bibinfo{author}{\bibfnamefont{R.~K.} \bibnamefont{Ellis}}, \bibnamefont{and}
  \bibinfo{author}{\bibfnamefont{C.}~\bibnamefont{Williams}},
  \bibinfo{journal}{JHEP} \textbf{\bibinfo{volume}{06}}, \bibinfo{pages}{179}
  (\bibinfo{year}{2016}), \eprint{1601.00658}.

\bibitem[{\citenamefont{Boughezal et~al.}(2017)\citenamefont{Boughezal,
  Campbell, Ellis, Focke, Giele, Liu, Petriello, and
  Williams}}]{Boughezal:2016wmq}
\bibinfo{author}{\bibfnamefont{R.}~\bibnamefont{Boughezal}},
  \bibinfo{author}{\bibfnamefont{J.~M.} \bibnamefont{Campbell}},
  \bibinfo{author}{\bibfnamefont{R.~K.} \bibnamefont{Ellis}},
  \bibinfo{author}{\bibfnamefont{C.}~\bibnamefont{Focke}},
  \bibinfo{author}{\bibfnamefont{W.}~\bibnamefont{Giele}},
  \bibinfo{author}{\bibfnamefont{X.}~\bibnamefont{Liu}},
  \bibinfo{author}{\bibfnamefont{F.}~\bibnamefont{Petriello}},
  \bibnamefont{and} \bibinfo{author}{\bibfnamefont{C.}~\bibnamefont{Williams}},
  \bibinfo{journal}{Eur. Phys. J.} \textbf{\bibinfo{volume}{C77}},
  \bibinfo{pages}{7} (\bibinfo{year}{2017}), \eprint{1605.08011}.

\bibitem[{\citenamefont{Assadsolimani et~al.}(2014)\citenamefont{Assadsolimani,
  Kant, Tausk, and Uwer}}]{Assadsolimani:2014oga}
\bibinfo{author}{\bibfnamefont{M.}~\bibnamefont{Assadsolimani}},
  \bibinfo{author}{\bibfnamefont{P.}~\bibnamefont{Kant}},
  \bibinfo{author}{\bibfnamefont{B.}~\bibnamefont{Tausk}}, \bibnamefont{and}
  \bibinfo{author}{\bibfnamefont{P.}~\bibnamefont{Uwer}},
  \bibinfo{journal}{Phys. Rev.} \textbf{\bibinfo{volume}{D90}},
  \bibinfo{pages}{114024} (\bibinfo{year}{2014}), \eprint{1409.3654}.

\bibitem[{\citenamefont{Meyer}(2017)}]{Meyer:2016slj}
\bibinfo{author}{\bibfnamefont{C.}~\bibnamefont{Meyer}},
  \bibinfo{journal}{JHEP} \textbf{\bibinfo{volume}{04}}, \bibinfo{pages}{006}
  (\bibinfo{year}{2017}), \eprint{1611.01087}.

\bibitem[{\citenamefont{Dulat et~al.}(2016)\citenamefont{Dulat, Hou, Gao,
  Guzzi, Huston, Nadolsky, Pumplin, Schmidt, Stump, and Yuan}}]{Dulat:2015mca}
\bibinfo{author}{\bibfnamefont{S.}~\bibnamefont{Dulat}},
  \bibinfo{author}{\bibfnamefont{T.-J.} \bibnamefont{Hou}},
  \bibinfo{author}{\bibfnamefont{J.}~\bibnamefont{Gao}},
  \bibinfo{author}{\bibfnamefont{M.}~\bibnamefont{Guzzi}},
  \bibinfo{author}{\bibfnamefont{J.}~\bibnamefont{Huston}},
  \bibinfo{author}{\bibfnamefont{P.}~\bibnamefont{Nadolsky}},
  \bibinfo{author}{\bibfnamefont{J.}~\bibnamefont{Pumplin}},
  \bibinfo{author}{\bibfnamefont{C.}~\bibnamefont{Schmidt}},
  \bibinfo{author}{\bibfnamefont{D.}~\bibnamefont{Stump}}, \bibnamefont{and}
  \bibinfo{author}{\bibfnamefont{C.~P.} \bibnamefont{Yuan}},
  \bibinfo{journal}{Phys. Rev.} \textbf{\bibinfo{volume}{D93}},
  \bibinfo{pages}{033006} (\bibinfo{year}{2016}), \eprint{1506.07443}.

\bibitem[{\citenamefont{Czakon et~al.}(2017)\citenamefont{Czakon, Heymes, and
  Mitov}}]{1606.03350}
\bibinfo{author}{\bibfnamefont{M.}~\bibnamefont{Czakon}},
  \bibinfo{author}{\bibfnamefont{D.}~\bibnamefont{Heymes}}, \bibnamefont{and}
  \bibinfo{author}{\bibfnamefont{A.}~\bibnamefont{Mitov}},
  \bibinfo{journal}{JHEP} \textbf{\bibinfo{volume}{04}}, \bibinfo{pages}{071}
  (\bibinfo{year}{2017}), \eprint{1606.03350}.

\bibitem[{\citenamefont{Gao et~al.}(2018)\citenamefont{Gao, Harland-Lang, and
  Rojo}}]{1709.04922}
\bibinfo{author}{\bibfnamefont{J.}~\bibnamefont{Gao}},
  \bibinfo{author}{\bibfnamefont{L.}~\bibnamefont{Harland-Lang}},
  \bibnamefont{and} \bibinfo{author}{\bibfnamefont{J.}~\bibnamefont{Rojo}},
  \bibinfo{journal}{Phys. Rept.} \textbf{\bibinfo{volume}{742}},
  \bibinfo{pages}{1} (\bibinfo{year}{2018}), \eprint{1709.04922}.

\bibitem[{\citenamefont{Bertone et~al.}(2018)\citenamefont{Bertone, Glazov,
  Mitov, Papanastasiou, and Ubiali}}]{Bertone:2017djs}
\bibinfo{author}{\bibfnamefont{V.}~\bibnamefont{Bertone}},
  \bibinfo{author}{\bibfnamefont{A.}~\bibnamefont{Glazov}},
  \bibinfo{author}{\bibfnamefont{A.}~\bibnamefont{Mitov}},
  \bibinfo{author}{\bibfnamefont{A.}~\bibnamefont{Papanastasiou}},
  \bibnamefont{and} \bibinfo{author}{\bibfnamefont{M.}~\bibnamefont{Ubiali}},
  \bibinfo{journal}{JHEP} \textbf{\bibinfo{volume}{04}}, \bibinfo{pages}{046}
  (\bibinfo{year}{2018}), \eprint{1711.03355}.

\bibitem[{\citenamefont{Collins}(1998)}]{Collins:1998rz}
\bibinfo{author}{\bibfnamefont{J.~C.} \bibnamefont{Collins}},
  \bibinfo{journal}{Phys. Rev.} \textbf{\bibinfo{volume}{D58}},
  \bibinfo{pages}{094002} (\bibinfo{year}{1998}), \eprint{hep-ph/9806259}.

\bibitem[{\citenamefont{Krämer et~al.}(2000)\citenamefont{Krämer, Olness, and
  Soper}}]{Kramer:2000hn}
\bibinfo{author}{\bibfnamefont{M.}~\bibnamefont{Krämer}},
  \bibinfo{author}{\bibfnamefont{F.~I.} \bibnamefont{Olness}},
  \bibnamefont{and} \bibinfo{author}{\bibfnamefont{D.~E.} \bibnamefont{Soper}},
  \bibinfo{journal}{Phys. Rev.} \textbf{\bibinfo{volume}{D62}},
  \bibinfo{pages}{096007} (\bibinfo{year}{2000}), \eprint{hep-ph/0003035}.

\bibitem[{\citenamefont{Abelleira~Fernandez
  et~al.}(2012)}]{AbelleiraFernandez:2012cc}
\bibinfo{author}{\bibfnamefont{J.}~\bibnamefont{Abelleira~Fernandez}}
  \bibnamefont{et~al.} (\bibinfo{collaboration}{LHeC Study Group}),
  \bibinfo{journal}{J.Phys.} \textbf{\bibinfo{volume}{G39}},
  \bibinfo{pages}{075001} (\bibinfo{year}{2012}), \eprint{1206.2913}.

\bibitem[{\citenamefont{Czakon et~al.}(2019)\citenamefont{Czakon, Dulat, Hou,
  Huston, Mitov, Papanastasiou, Sitiwaldi, Yu, and Yuan}}]{1912.08801}
\bibinfo{author}{\bibfnamefont{M.}~\bibnamefont{Czakon}},
  \bibinfo{author}{\bibfnamefont{S.}~\bibnamefont{Dulat}},
  \bibinfo{author}{\bibfnamefont{T.-J.} \bibnamefont{Hou}},
  \bibinfo{author}{\bibfnamefont{J.}~\bibnamefont{Huston}},
  \bibinfo{author}{\bibfnamefont{A.}~\bibnamefont{Mitov}},
  \bibinfo{author}{\bibfnamefont{A.~S.} \bibnamefont{Papanastasiou}},
  \bibinfo{author}{\bibfnamefont{I.}~\bibnamefont{Sitiwaldi}},
  \bibinfo{author}{\bibfnamefont{Z.}~\bibnamefont{Yu}}, \bibnamefont{and}
  \bibinfo{author}{\bibfnamefont{C.~P.} \bibnamefont{Yuan}}
  (\bibinfo{year}{2019}), \eprint{1912.08801}.

\bibitem[{\citenamefont{Gao et~al.}(2013)\citenamefont{Gao, Li, and
  Zhu}}]{1210.2808}
\bibinfo{author}{\bibfnamefont{J.}~\bibnamefont{Gao}},
  \bibinfo{author}{\bibfnamefont{C.~S.} \bibnamefont{Li}}, \bibnamefont{and}
  \bibinfo{author}{\bibfnamefont{H.~X.} \bibnamefont{Zhu}},
  \bibinfo{journal}{Phys. Rev. Lett.} \textbf{\bibinfo{volume}{110}},
  \bibinfo{pages}{042001} (\bibinfo{year}{2013}), \eprint{1210.2808}.

\bibitem[{\citenamefont{Gao and Papanastasiou}(2017)}]{Gao:2017goi}
\bibinfo{author}{\bibfnamefont{J.}~\bibnamefont{Gao}} \bibnamefont{and}
  \bibinfo{author}{\bibfnamefont{A.~S.} \bibnamefont{Papanastasiou}},
  \bibinfo{journal}{Phys. Rev.} \textbf{\bibinfo{volume}{D96}},
  \bibinfo{pages}{051501} (\bibinfo{year}{2017}), \eprint{1705.08903}.

\bibitem[{\citenamefont{Liu and Gao}(2018)}]{Liu:2018gxa}
\bibinfo{author}{\bibfnamefont{Z.~L.} \bibnamefont{Liu}} \bibnamefont{and}
  \bibinfo{author}{\bibfnamefont{J.}~\bibnamefont{Gao}},
  \bibinfo{journal}{Phys. Rev.} \textbf{\bibinfo{volume}{D98}},
  \bibinfo{pages}{071501} (\bibinfo{year}{2018}), \eprint{1807.03835}.

\bibitem[{\citenamefont{Behring et~al.}(2019)\citenamefont{Behring, Czakon,
  Mitov, Papanastasiou, and Poncelet}}]{Behring:2019iiv}
\bibinfo{author}{\bibfnamefont{A.}~\bibnamefont{Behring}},
  \bibinfo{author}{\bibfnamefont{M.}~\bibnamefont{Czakon}},
  \bibinfo{author}{\bibfnamefont{A.}~\bibnamefont{Mitov}},
  \bibinfo{author}{\bibfnamefont{A.~S.} \bibnamefont{Papanastasiou}},
  \bibnamefont{and} \bibinfo{author}{\bibfnamefont{R.}~\bibnamefont{Poncelet}},
  \bibinfo{journal}{Phys. Rev. Lett.} \textbf{\bibinfo{volume}{123}},
  \bibinfo{pages}{082001} (\bibinfo{year}{2019}), \eprint{1901.05407}.

\bibitem[{\citenamefont{Sirunyan et~al.}(2020{\natexlab{b}})}]{1907.08330}
\bibinfo{author}{\bibfnamefont{A.~M.} \bibnamefont{Sirunyan}}
  \bibnamefont{et~al.} (\bibinfo{collaboration}{CMS}), \bibinfo{journal}{Eur.
  Phys. J.} \textbf{\bibinfo{volume}{C80}}, \bibinfo{pages}{370}
  (\bibinfo{year}{2020}{\natexlab{b}}), \eprint{1907.08330}.

\bibitem[{\citenamefont{Aaboud et~al.}(2017{\natexlab{b}})}]{Aaboud:2016ymp}
\bibinfo{author}{\bibfnamefont{M.}~\bibnamefont{Aaboud}} \bibnamefont{et~al.}
  (\bibinfo{collaboration}{ATLAS}), \bibinfo{journal}{JHEP}
  \textbf{\bibinfo{volume}{04}}, \bibinfo{pages}{086}
  (\bibinfo{year}{2017}{\natexlab{b}}), \eprint{1609.03920}.

\bibitem[{\citenamefont{Sirunyan
  et~al.}(2017{\natexlab{b}})}]{Sirunyan:2016cdg}
\bibinfo{author}{\bibfnamefont{A.~M.} \bibnamefont{Sirunyan}}
  \bibnamefont{et~al.} (\bibinfo{collaboration}{CMS}), \bibinfo{journal}{Phys.\
  Lett.\ B} \textbf{\bibinfo{volume}{772}}, \bibinfo{pages}{752}
  (\bibinfo{year}{2017}{\natexlab{b}}), \eprint{1610.00678}.

\end{thebibliography}

\end{document}